
This paper is typeset with AMSTeX 2.1.  If you do not have this addition
to TeX, contact the TeX User Group at
	TeX Users Group
	P.O. Box 869
	Santa Barbara, CA 93102
	USA
	(805)963-1338
	(805)963-8358 (fax)
or by anonymous ftp to e-math.ams.org (the creators of AMSTeX).

\documentstyle{amsppt}

\expandafter\redefine\csname logo\string @\endcsname{}

\NoBlackBoxes
\magnification=\magstep1
\hsize=6.5truein
\vsize=8.5truein
\topmatter
\title
  Some quantum analogues of solvable Lie groups
  \endtitle \author C. De Concini, V.G. Kac, and C. Procesi
  \endauthor
  \endtopmatter \document \baselineskip=.15truein

\subheading{Introduction}

In the papers [DK1-2],[DKP1-2] the quantized enveloping algebras
introduced by Drinfeld and Jimbo have been studied in the case $q =
\varepsilon$, a primitive $l$-th root of $1$ with $l$ odd (cf. \S 2
for basic definitions).  Let us only recall for the moment that such
algebras are canonically constructed starting from a Cartan matrix of
finite type and in particular we can talk of the associated classical
objects (the root system, the simply connected algebraic group $G$,
etc.).  For such an algebra the generic (resp. any) irreducible
representation has dimension equal to (resp. bounded by) $l^{N}$
where $N$ is the number of positive roots and the set of irreducible
representations has a canonical map  to the big cell of the
corresponding group $G$.

In this paper we analyze the structure of some subalgebras of quantized
enveloping algebras
corresponding to unipotent and solvable subgroups of $G$.  These algebras
have the non-commutative structure of iterated algebras of twisted
polynomials with a derivation, an object which has often appeared in
the general theory of non-commutative rings (see e.g. [KP], [GL] and
references there).  In particular, we find maximal dimensions of
their irreducible representations.  Our results confirm the
validity of the general philosophy that the representation theory is
intimately connected to the Poisson geometry.

\subheading{\S 1. Twisted polynomial rings}

{\bf 1.1.}  In this section we will collect some well known
definitions and properties of twisted derivations.

Let $A$ be an algebra and let $\sigma$ be an automorphism of $A$.
A {\it twisted derivation} of $A$ relative to
$\sigma$ is a linear map $D:\ A @>>> A$ such that:

$$D(ab) = D(a)b + \sigma (a)D(b).$$

\vskip 10pt
\noindent {\bf Example.} An element $a \in A$ induces an inner
twisted derivation $ad_{\sigma}a$ relative to $\sigma$ defined by
the formula:

$$(ad_{\sigma}a)b = ab - \sigma (b)a.$$

The following well-known fact is very useful in
calculations with twisted derivations.  (Here and further we use the
usual ``box'' notation:

$$[n] = \frac{q^{n}-q^{-n}}{q-q^{-1}},\ [n]! = [1][2]\ldots [n],\
\bmatrix m \\ n\endbmatrix = \frac{[m][m-1]\ldots
  [m-n+1]}{[n]!}$$ One also writes $[n]_{d}$, etc. if $q$ is
replaced by $q^{d}$.)

\proclaim{Proposition} Let $a \in A$ and let $\sigma$ be an
automorphism of $A$ such that $\sigma (a) = q^{2}a$, where $q$ is a
scalar.  Then

$$(ad_{\sigma}a)^{m}(x) = \sum^{m}_{j=0} (-1)^{j} q^{j(m-1)}
\bmatrix m \\ j\endbmatrix a^{m-j}\sigma^{j}(x)a^{j}.$$

\endproclaim

\demo{Proof} Let $L_{a}$ and $R_{a}$ denote the operators of left and
right multiplications by $a$ in $A$.  Then

$$ad_{\sigma}a = L_{a} - R_{a}\sigma.$$
Since $L_{a}$ and $R_{a}$ commute, due to the assumption $\sigma (a)
= q^{2}a$ we have

$$L_{a}(R_{a}\sigma ) = q^{-2}(R_{a}\sigma )L_{a}.$$
Now the proposition is immediate from the following well-known
binomial formula applied to the algebra End $A$.

\enddemo

\proclaim{Lemma}  Suppose that $x$ and $y$ are elements of an algebra
such that $yx = q^{2}xy$ for some scalar $q$.  Then

$$(x+y)^{m} = \sum^{m}_{j=0} \bmatrix m \\ j\endbmatrix
q^{j(m-j)} x^{j} y^{m-j}.$$
\endproclaim

\demo\nofrills{Proof}\ is by induction on $m$ using

$$\bmatrix m \\ j-1\endbmatrix q^{m+1} + \bmatrix m \\
j\endbmatrix = \bmatrix m+1 \\ j\endbmatrix q^{j},$$ which
follows from

$$q^{b}[a] + q^{-a}[b] = [a+b].\ \ \square$$
\enddemo

Let $\ell$ be a positive integer and let $q$ be a primitive
$\ell$-th root of $1$.  Let $\ell^{\prime} = \ell$ if $\ell$ is odd
and $= \frac{1}{2} \ell$ if $\ell$ is even.  Then, by definition, we
have
$$\bmatrix \ell^{\prime} \\ j \endbmatrix = 0 \ \text{for all}\
j\ \text{such that}\ 0 < j < \ell^{\prime}.$$ This together with
Proposition 1.1 implies

\proclaim{Corollary} Under the hypothesis of Proposition 1.1 we have:

$$(ad_{\sigma}a)^{\ell^{\prime}} (x) = a^{\ell^{\prime}}x -
\sigma^{\ell^{\prime}}(x)a^{\ell^{\prime}}\ \text{if}\ q\ \text{is a
primitive}\ \ell\text{-th root of}\ 1.$$
\endproclaim

\noindent {\bf Remark.} Let $D$ be a twisted derivation
associated to an automorphism $\sigma$ such that $\sigma D =
q^{2}D\sigma$.  Then by induction on $m$ one obtains the
following well-known $q$-analogue of the Leibnitz formula:

$$D^{m}(xy) = \sum^{m}_{j=0} \bmatrix m \\ j \endbmatrix
q^{j(m-j)} D^{m-j}(\sigma^{j}x)D^{j}(y).$$ It follows that if $q$
is a primitive $\ell$-th root of $1$, then $D^{\ell^{\prime}}$ is
a twisted derivation associated to $\sigma^{\ell^{\prime}}$.

{\bf 1.2.} Given an automorphism $\sigma$ of $A$ and a twisted
derivation $D$ of $A$ relative to $\sigma$ we define the {\it twisted
polynomial algebra} $A_{\sigma,D}[x]$ in the indeterminate $x$ to be
the ${\Bbb F}$-module $A \otimes_{\Bbb F} {\Bbb F}[x]$ thought as
formal polynomials with multiplication defined by the rule:

$$xa = \sigma(a)x + D(a).$$
When $D = 0$ we will also denote this ring by $A_{\sigma}[x]$.
Notice that the definition has been chosen in such a way that in the
new ring the given twisted derivation becomes the inner derivation
$ad_{\sigma}x$.

Let us notice that if $a,b \in A$ and $a$ is invertible we can
perform the change of variables $y := ax + b$ and we see that
$A_{\sigma ,D}[x] = A_{\sigma^{\prime},D^{\prime}}[y]$.  It is better
to make the formulas explicit separately when $b = 0$ and when $a =
1$.  In the first case $yc = axc = a(\sigma (c)x + D(c)) = a(\sigma
(c))a^{-1}y + aD(c)$ and we see that the new automorphism
$\sigma^{\prime}$ is the composition $(Ada)\sigma$, so that
$D^{\prime} := aD$ is a twisted derivation relative to
$\sigma^{\prime}$.  Here and further $Ada$ stands for the inner
automorphism:

$$(Ada)x = axa^{-1}.$$
In the case $a = 1$ we have $yc = (x + b)c = \sigma (c)x + D(b) + bc
= \sigma (c)y + D(b) + bc - \sigma (c)b$, so that $D^{\prime} = D +
ad_{\sigma}b$.  Summarizing we have

\proclaim{Proposition} Changing $\sigma ,D$ to $(Ada)\sigma ,aD$
(resp. to $\sigma ,D + D_{b}$) does not change the twisted polynomial
ring up to isomorphism.\ \ \ $\square$
\endproclaim

We may express the previous fact with a definition: For a ring
$A$ two pairs $(\sigma ,D)$ and $(\sigma^{\prime},D^{\prime})$
are {\it equivalent} if they are obtained one from the other by
the above moves.

If $D = 0$ we can also consider the twisted Laurent polynomial
algebra $A_{\sigma}[x,x^{-1}]$.  It is clear that if $A$ has no zero
divisors, then the algebras $A_{\sigma ,D}[x]$ and
$A_{\sigma}[x,x^{-1}]$ also have no zero divisors.

The importance for us of twisted polynomial algebras will be clear in
the section on quantum groups.

{\bf 1.3.}  We want to study special cases of the previous
construction.

Let us first consider a finite dimensional semisimple algebra $A$
over an algebraically closed field ${\Bbb F}$, let
$\underset{i}\to{\oplus} {\Bbb F} e_{i}$ be the fixed points of the
center of $A$ under $\sigma$ where the $e_{i}$ are central
idempotents.  We have $D(e_{i}) = D(e^{2}_{i}) = 2D(e_{i})e_{i}$
hence $D(e_{i}) = 0$ and, if $x = xe_{i}$, then $D(x) = D(x)e_{i}$.  It
follows that, decomposing $A = \underset{i}\to{\oplus} Ae_{i}$, each
component $Ae_{i}$ is stable under $\sigma$ and $D$ and thus we have

$$A_{\sigma ,D}[x] = \underset{i}\to{\oplus} (Ae_{i})_{\sigma
,D}[x].$$
This allows us to restrict our analysis to the case in which $1$ is
the only fixed central idempotent.

The second special case is described by the following:

\proclaim{Lemma} Consider the algebra $A = {\Bbb F}^{\oplus k}$ with
$\sigma$ the cyclic permutation of the summands, and let $D$ be a
twisted derivation of this algebra relative to $\sigma$.  Then $D$ is
an inner twisted derivation.
\endproclaim

\demo{Proof} Compute $D$ on the idempotents: $D(e_{i}) =
D(e^{2}_{i}) = D(e_{i})(e_{i} + e_{i+1})$.  Hence we must have $D(e_{i}) =
a_{i}e_{i} - b_{i}e_{i+1}$ and from $0 = D(e_{i}e_{i+1}) =
D(e_{i})e_{i+1} + e_{i+1}D(e_{i+1})$ we deduce $b_{i} = a_{i+1}$.
Let now $a = (a_{1},a_{2},\ldots ,a_{k})$; an easy computation shows
that $D = ad_{\sigma}a$.\ \ \ $\square$
\enddemo

\proclaim{Proposition} Let $\sigma$ be the cyclic permutation of the
summands of the algebra ${\Bbb F}^{\oplus k}$.  Then

(a) ${\Bbb F}^{\oplus k}_{\sigma}[x,x^{-1}]$ is an Azumaya algebra of
degree $k$ over its center ${\Bbb F}[x^{k},x^{-k}]$.

(b) $R := {\Bbb F}^{\oplus k}_{\sigma} [x,x^{-1}] \otimes_{{\Bbb
F}[x^{k},x^{-k}]} {\Bbb F}[x,x^{-1}]$ is the algebra of $k \times k$
matrices over ${\Bbb F}[x,x^{-1}]$.
\endproclaim

\demo{Proof} It is enough to prove (b).  Let $u := x \otimes x^{-1},\
e_{i} := e_{i} \otimes 1$; we have $u^{k} = x^{k} \otimes x^{-k}
= 1$ and $ue_{i} = e_{i+1}u$.  From these formulas it easily follows
that the elements $e_{i}u^{j}\ (i,j = 1,\ldots,k)$ span a subalgebra
$A$ and that there exists an isomorphism $A
\widetilde{\longrightarrow} M_{k}({\Bbb F})$ mapping ${\Bbb
F}^{\oplus k}$ to the diagonal matrices and
$u$ to the matrix of the cyclic permutation.  Then $R = A
\otimes_{\Bbb F} {\Bbb F}[x,x^{-1}]$.\ \ \ $\square$
\enddemo

{\bf 1.4.}  Assume now that $A$ is semisimple and that $\sigma$
induces a cyclic permutation of the central idempotents.

\proclaim{Lemma} (a) $A = M_{d}({\Bbb F})^{\oplus k}$.

(b) Let $D$ be a twisted derivation of $A$ relative to $\sigma$.
Then the pair $(\sigma ,D)$ is equivalent to the pair
$(\sigma^{\prime},0)$ where

$$\sigma^{\prime}(a_{1},a_{2},\ldots ,a_{k}) =
(a_{k},a_{1},a_{2},\ldots ,a_{k-1}). \tag{1.4.1}$$
\endproclaim

\demo{Proof} Since $\sigma$ permutes transitively the simple blocks
they must all have the same degree $d$ so that $A = M_{d}(F)^{\oplus
k}$.  Furthermore we can arrange the identifications of the simple
blocks with matrices so that:

$$\sigma (a_{1},a_{2},\ldots ,a_{k}) = (\tau
(a_{k}),a_{1},a_{2},\ldots ,a_{k-1}),$$
where $\tau$ is an automorphism of $M_{d}({\Bbb F})$.  Any such
automorphism is inner, hence after composing $\sigma$ with an inner
automorphism, we may assume in the previous formula that $\tau = 1$.
Then we think of $A$ as $M_{d}({\Bbb F}) \otimes {\Bbb F}^{\oplus
k}$, the new automorphism being of the form $1 \otimes
\sigma^{\prime}$ where $\sigma^{\prime} :\ F^{\oplus k} @>>>
F^{\oplus k}$ is given by (1.4.1).

We also have that $M_{d}({\Bbb F}) = A^{\sigma}$ and ${\Bbb
F}^{\oplus k}$ is the centralizer of $A^{\sigma}$.  Next observe that
$D$ restricted to $A^{\sigma}$ is a derivation of $M_{d}({\Bbb F})$
with values in $\oplus^{k}_{i=1} M_{d}({\Bbb F})$, i.e., $D(a) =
(D_{1}(a),D_{2}(a),\ldots ,D_{k}(a))$ where each $D_{i}$ is a
derivation of $M_{d}({\Bbb F})$.  Since for $M_{d}({\Bbb F})$ all
derivations are inner we can find an element $u \in A$ such that
$D(a) = [u,a]$ for all $a \in M_{d}({\Bbb F})$.  So $(D -
ad_{\sigma}u)(a) = [u,a] - (ua - \sigma (a)u) = 0$ for $a \in
A^{\sigma}$.  Thus, changing $D$ by adding $-ad_{\sigma}u$ we may
assume that $D = 0$ on $M_{d}({\Bbb F})$.

Now consider $b \in {\Bbb F}^{\oplus k}$ and $a \in M_{d}({\Bbb F})$;
we have $D(b)a = D(ba) = D(ab) = aD(b)$.  Since ${\Bbb F}^{\oplus k}$
is the centralizer of $M_{d}({\Bbb F})$ we have $D(b) \in {\Bbb
F}^{\oplus k}$ and $D$ induces a twisted derivation of ${\Bbb
F}^{\oplus k}$.  By Lemma 1.3 this last derivation is inner and the
claim is proved.\ \ \ $\square$
\enddemo

Summarizing we have

\proclaim{Proposition}  Let $A$ be a finite-dimensional semisimple
algebra over an algebraically closed field ${\Bbb F}$.  Let $\sigma$
be an automorphism of $A$ which induces a cyclic permutation of order
$k$ of the central idempotents of $A$.  Let $D$ be a twisted
derivation of $A$ relative to $\sigma$.  Then:

$$\align
A_{\sigma ,D}[x] &\cong M_{d}({\Bbb F}) \otimes {\Bbb F}^{\oplus
k}_{\sigma}[x], \\
A_{\sigma ,D}[x,x^{-1}] &\cong M_{d}({\Bbb F}) \otimes {\Bbb
F}^{\oplus k}_{\sigma}[x,x^{-1}].
\endalign
$$
This last algebra is Azumaya of degree $dk$.
\endproclaim

{\bf 1.5.} We can now globalize the previous construction.  Let $A$
be a prime algebra (i.e. $aAb = 0,\ a,b \in A$, implies that $a= 0$ or $b =
0$) over a field ${\Bbb F}$ and let $Z$ be the center of $A$.  Then $Z$ is a
domain and $A$ is a torsion free module over $Z$.  Assume that $A$ is a finite
module over $Z$.  Then $A$ embeds in a finite-dimensional central simple
algebra $Q(A) = A
\otimes_{Z} Q(Z)$, where $Q(Z)$ is the
ring of fractions of $Z$.  If $\overline{Q(Z)}$ denotes the algebraic
closure of $Q(Z)$ we have that $A \otimes_{Z} \overline{Q(Z)}$ is the
full ring $M_{d}(\overline{Q(Z)})$ of $d \times d$ matrices over
$\overline{Q(Z)}$.  Then $d$ is called the {\it degree} of $A$.

Let $\sigma$ be an automorphism of the algebra $A$ and let $D$ be a
twisted derivation of $A$ relative to $\sigma$.  Assume that

(a) There is a subalgebra $Z_{0}$ of $Z$, such that $Z$ is finite
over $Z_{0}$.

(b) $D$ vanishes on $Z_{0}$ and $\sigma$ restricted to $Z_{0}$ is the
identity.\newline
\noindent These assumptions imply that $\sigma$ restricted to $Z$
is an automorphism of finite order.  Let $d$ be the degree of $A$
and let $k$ be the order of $\sigma$ on the center $Z$.  Assume
that the field ${\Bbb F}$ has characteristic $0$.  The main
result of this section is:

\proclaim{Theorem} Under the above assumptions the twisted polynomial
 algebra $A_{\sigma ,D}[x]$ is an order in a central simple algebra
of degree $kd$.
\endproclaim

\demo{Proof}
Let $Z^{\sigma}$ be the fixed points in $Z$ of $\sigma$.
By the definition, it is clear that $D$ restricted to $Z^{\sigma}$
is a derivation.  Since it vanishes on a subalgebra over which it is
finite hence algebraic and since we are in characteristic zero it
follows that $D$ vanishes on $Z^{\sigma}$.  Let us embed $Z^{\sigma}$
in an algebraically closed field $L\/ $ and let us consider the
algebra $A \otimes_{Z^{\sigma}}L\/$.  This algebra of course
equals $A \otimes_{Z} Z \otimes_{Z^{\sigma}} L$, but clearly $Z
\otimes_{Z^{\sigma}} L\/ = L\/^{\oplus k}$ and $A
\otimes_{Z} L\/ = M_{d}(L\/)$.  Thus we get that $A
\otimes_{Z^{\sigma}}L\/ = \oplus^{k}_{i=1} M_{d}(L\/)$.
The pair $\sigma ,D$ extends to $A \otimes_{Z^{\sigma}} L\/$ and
using the same notations we have that $(A \otimes_{Z^{\sigma}} L
)_{\sigma ,D}[x] = (A_{\sigma ,D}[x]) \otimes_{Z^{\sigma}} L$.
We are now in the situation of a semisimple algebra which we
have already studied and the claim follows.\ \ \ $\square$
\enddemo

\proclaim{Corollary} Under the above assumptions, $A_{\sigma
,D}[x]$ and $A_{\sigma}[x]$ have the same degree.
\endproclaim

\noindent {\bf Remark.} The previous analysis yeilds in fact a
stronger result.  Consider the open set of $\text{Spec}\ Z$ where
$A$ is an Azumaya algebra; it is clearly $\sigma$-stable.  In it
we consider the open part where $\sigma$ has order exactly $k$.
Every orbit of $k$ elements of the group generated by $\sigma$
gives a point $F(p)$ in $\text{Spec}\ Z^{\sigma}$ and $A
\otimes_{Z} Z \otimes_{Z^{\sigma}} F(p) = \oplus^{k}_{i=1}
M_{d}(F(p))$.  Thus we can apply the previous theory which allows
us to describe the fiber over $F(p)$ of the spectrum of
$A_{\sigma ,D}[x]$.

{\bf 1.6.} Let $A$ be a prime algebra over a field ${\Bbb F}$ of
characteristic $0$, let $x_{1},\ldots ,x_{n}$ be a set of
generators of $A$ and let $Z_{0}$ be a central subalgebra of $A$.
For each $i = 1,\ldots ,k$, denote by $A^{i}$ the subalgebra of
$A$ generated by $x_{1},\ldots ,x_{i}$, and let $Z^{i}_{0} =
Z_{0} \cap A^{i}$.  We assume that the following three conditions
hold for each $i = 1,\ldots ,k$:

(a) $x_{i}x_{j} = b_{ij}x_{j}x_{i} + P_{ij}$ if $i > j$, where
$b_{ij} \in {\Bbb F},\ P_{ij} \in A^{i-1}$.

(b) $A^{i}$ is a finite module over $Z^{i}_{0}$.

(c) Formulas $\sigma_{i}(x_{j}) = b_{ij}x_{j}$ for $j < i$ define an
automorphism of $A^{i-1}$ which is the identity on $Z^{i-1}_{0}$.

Note that letting $D_{i}(x_{j}) = P_{ij}$ for $j < i$, we obtain
$A^{i} = A^{i-1}_{\sigma_{i},D_{i}}[x_{i}],$ so that $A$ is an
iteratated twisted polynomial algebra.  Note also that each
triple $(A^{i-1},\sigma_{i},D_{i})$ satisfies assumptions 1.5(a)
and (b).

We may consider the twisted polynomial algebras $\overline{A}^{i}$
with zero derivations, so that the relations are $x_{i}x_{j} =
b_{ij}x_{j}x_{i}$ for $j < i$.  We call this the {\it associated}
{\it quasipolynomial} {\it algebra} (as in [DK1]).

We can prove now the main theorem of this section.

\proclaim{Theorem} Under the above assumptions, the
degree of $A$ is equal to the degree of the associated
quasipolynomial algebra $\overline{A}$.
\endproclaim

\demo{Proof} We use this following remark.  If there is an index $h$
such that the elements $P_{ij} = 0$ for all $i > h$ and all $j$, then
the monomials in the variables different from $x_{h}$ form a
subalgebra $B$ and the algebra $A$ is a twisted polynomial ring
$B_{\sigma ,D}[x_{h}]$.  The associated ring $B_{\sigma}[x_{h}]$ is
obtained by setting $P_{hj} = 0$ for all $j$.  Having made this
remark we see that we can inductively modify the relations 1.6(a) so
that at the $h$-th step we have an algebra $A^{n}_{h}$ with the same
type of relations but $P_{ij} = 0$ for all $i > n - h$ and all $j$.
Since $A^{n}_{h}$ and $A^{n}_{h-1}$ are of type $B_{\sigma ,D}[x]$
and $B_{\sigma}[x]$ respectively we see, by Corollary 1.5, that they
have all the same degree.\ \ \ $\square$
\enddemo

\subheading{\S 2. Quantum groups}

{\bf 2.1.} Let $(a_{ij})$ be an indecomposable $n \times n$ Cartan
matrix and let $d_{1},\ldots ,d_{n}$ be relatively prime positive
integers such that $d_{i}a_{ij} = d_{j}a_{ji}$.  Recall the
associated notions of the weight, coroot and root lattices
$P,Q^{\vee}$ and $Q$, of the root and coroot systems $R$ and
$R^{\vee}$, of the Weyl group $W$, the $W$-invariant bilinear form
$(.|.)$, etc.:

Let $P$ be a lattice over ${\Bbb Z}$ with basis $\omega_{1},\ldots
,\omega_{n}$ and let $Q^{\vee} = \text{Hom}_{\Bbb Z}(P,{\Bbb Z})$ be
the dual lattice with the dual basis $\alpha^{\vee}_{1},\ldots
,\alpha^{\vee}_{n}$, i.e. $\langle
\omega_{i},\alpha^{\vee}_{j}\rangle = \delta_{ij}$.  Let $P_{+} =
\sum^{n}_{i=1}{\Bbb Z}_{+}\omega_{i}$.  Let

$$\rho = \sum^{n}_{i=1} \omega_{i},\ \alpha_{j} = \sum^{n}_{i=1}
a_{ij}\omega_{i}\ (j = 1,\ldots ,n),$$
and let $Q = \sum^{n}_{j=1} {\Bbb Z}\alpha_{j} \subset P$, and $Q_{+}
= \sum^{n}_{j=1} {\Bbb Z}_{+} \alpha_{j}$.

Define automorphisms $s_{i}$ of $P$ by $s_{i}(\omega_{j}) =
\omega_{j} - \delta_{ij}\alpha_{i}\ (i,j = 1,\ldots ,n)$.  Then
$s_{i}(\alpha_{j}) = \alpha_{j} - a_{ij}\alpha_{i}$.  Let $W$ be the
subgroup of $GL(P)$ generated by $s_{1},\ldots ,s_{n}$.  Let

$$\align
\Pi &= \{\alpha_{1},\ldots ,\alpha_{n}\},\ \Pi^{\vee} =
\{\alpha^{\vee}_{1},\ldots ,\alpha^{\vee}_{n}\}, \\
R &= W\Pi,\ R^{+} = R \cap Q_{+},\ R^{\vee} = W\Pi^{\vee}.
\endalign$$ The map $\alpha_{i} \longmapsto \alpha^{\vee}_{i}$
extends uniquely to a bijective $W$-equivariant map $\alpha
\longmapsto \alpha^{\vee}$ between $R$ and $R^{\vee}$.  The
reflection $s_{\alpha}$ defined by $s_{\alpha}(\lambda ) =
\lambda - \langle \lambda ,\alpha^{\vee}\rangle\alpha$ lies in
$W$ for each $\alpha \in R$, so that $s_{\alpha_{i}} = s_{i}$.

Define a bilinear pairing $P \times Q @>>> {\Bbb Z}$ by
$(\omega_{i}|\alpha_{j}) = \delta_{ij}d_{j}$.  Then
$(\alpha_{i}|\alpha_{j}) = d_{i}a_{ij}$, giving a symmetric ${\Bbb
Z}$-valued $W$-invariant bilinear form on $Q$ such that $(\alpha
|\alpha) \in 2{\Bbb Z}$.  We may identify $Q^{\vee}$ with a
sublattice of the ${\Bbb Q}$-span of $P$ (containing $Q$) using this
form.  Then:

$$\alpha^{\vee}_{i} = d^{-1}_{i}\alpha_{i},\ \alpha^{\vee} = 2\alpha
/(\alpha |\alpha ).$$

One defines the {\it simply} {\it connected quantum group} ${\Cal U}$
associated to the matrix $(a_{ij})$ as an algebra over the ring
${\Cal A} := {\Bbb C}[q,q^{-1},(q^{d_{i}} - q^{-d_{i}})^{-1}]$ on
generators $E_{i},F_{i}\ (i = 1,\ldots ,n),\ K_{\alpha}\  (\alpha \in
P)$ subject to the following relations (this is a simple variation of
the construction of Drinfeld and Jimbo):

$$\align
&K_{\alpha}K_{\beta} = K_{\alpha + \beta},\ K_{0} = 1,\\
&\sigma_{\alpha}(E_{i}) = q^{(\alpha |\alpha_{i})}E_{i},\
\sigma_{\alpha}(F_{i}) = q^{-(\alpha |\alpha_{i})} F_{i},\\
&[E_{i},F_{j}] = \delta_{ij}
\frac{K_{\alpha_{i}}-K_{-\alpha_{i}}}{q^{d_{i}}-q^{-d_{i}}}, \\
&(ad_{\sigma_{-\alpha_{i}}} E_{i})^{1-a_{ij}} E_{j} = 0,\
(ad_{\sigma_{-\alpha_{i}}} F_{i})^{1-a_{ij}} F_{j} = 0\ (i \neq j),
\endalign$$ where $\sigma_{\alpha} = Ad\ K_{\alpha}$.  Recall
that ${\Cal U}$ has a Hopf algebra structure with
comultiplication $\Delta$, antipode $S$ and counit $\eta$ defined
by:

$$\align
&\Delta E_{i} = E_{i} \otimes 1 + K_{\alpha_{i}} \otimes E_{i},\
\Delta F_{i} = F_{i} \otimes K_{-\alpha_{i}} + 1 \otimes F_{i},\
\Delta K_{\alpha} = K_{\alpha} \otimes K_{\alpha} , \\
&SE_{i} = -K_{-\alpha_{i}} E_{i},\ SF_{i} = -F_{i}K_{i},\ SK_{\alpha}
= K_{-\alpha}, \\
&\eta E_{i} = 0,\ \eta F_{i} = 0,\ \eta K_{\alpha} = 1.
\endalign$$

Recall that the braid group associated to $W$, whose
canonical generators one denotes by $T_{i}$, acts as a group of
automorphisms of the algebra ${\Cal U}$ ([L]):

$$\align
T_{i}K_{\alpha} &= K_{s_{i}(\alpha )},\ T_{i}E_{i} =
-F_{i}K_{\alpha_{i}}, \\
T_{i}E_{j} &= \frac{1}{[-a_{ij}]_{d_{i}}!}
(ad_{\sigma_{-\alpha_{i}}}(-E_{i}))^{-a_{ij}} E_{j}, \\
T_{i}\kappa &= \kappa T_{i},
\endalign$$ where $\kappa$ is a conjugate-linear
anti-automorphism of ${\Cal U}$, viewed as an algebra over ${\Bbb
  C}$, defined by:

$$\kappa E_{i} = F_{i},\ \kappa F_{i} = E_{i},\ \kappa K_{\alpha} =
K_{-\alpha},\ \kappa q = q^{-1}.$$

{\bf 2.2.}  Fix a reduced expression $w_{0} = s_{i_{1}} \ldots
s_{i_{N}}$ of the longest element of $W$, and let

$$\beta_{1} = \alpha_{i_{1}},\ \beta_{2} =
s_{i_{1}}(\alpha_{i_{2}}),\ldots ,\beta_{N} = s_{i_{1}}\ldots
s_{i_{N-1}}(\alpha_{i_{N}})$$
be the corresponding convex ordering of $R^{+}$.  Introduce the
corresponding {\it root vectors} $(m = 1,\ldots ,N)$ ([L]):

$$E_{\beta_{m}} = T_{i_{1}}\ldots T_{i_{m-1}}E_{i_{m}},\
E_{-\beta_{m}} = T_{i_{1}} \ldots T_{i_{m-1}}F_{i_{m}} = \kappa
E_{\beta_m}$$
(they depend on the choice of the reduced expression).

For $k = (k_{1},\ldots ,k_{N}) \in {\Bbb Z}^{N}_{+}$ we let

$$E^{k} = E^{k_{1}}_{\beta_{1}} \ldots E^{k_{N}}_{\beta_{N}},\ F^{k}
= \kappa E^{k}.$$

\proclaim{Lemma} (a) [L] The elements $F^{k}K_{\alpha}E^{r}$, where
$k,r \in {\Bbb Z}^{N}_{+},\ \alpha \in P$, form a basis of ${\Cal U}$
over ${\Cal A}$.

(b) [LS] For $i < j$ one has:

$$E_{\beta_{i}}E_{\beta_{j}} -
q^{(\beta_{i}|\beta_{j})}E_{\beta_{j}}E_{\beta_{i}} = \sum_{k \in
{\Bbb Z}^{N}_{+}} c_{k}E^{k}, \tag{2.2.1}$$
where $c_{k} \in {\Bbb C}[q,q^{-1}]$ and $c_{k} \neq 0$ only when $k
= (k_{1},\ldots ,k_{N})$ is such that $k_{s} = 0$ for $s \leq i$ and
$s \geq j$.\ \ \ $\square$
\endproclaim

An immediate corollary is the following:

Let $w$ be any element of the Weyl group.  We can choose for it a
reduced expression $w = s_{i_{1}} \ldots s_{i_{k}}$ which we complete
to a reduced expression $w_{0} = s_{i_{1}}\ldots s_{i_{N}}$ of the
longest element of $W$.  Consider the elements $E_{\beta_{j}},\ j =
1,\ldots ,k$.  Then we have:

\proclaim{Proposition} (a) The elements $E_{\beta_{j}},\ j = 1,\ldots
,k$, generate a subalgebra ${\Cal U}^{w}$ which is independent of the
choice of the reduced expression of $w$.

(b) If $w^{\prime} = ws$ with $s$ a simple reflection and
$l(w^{\prime}) = l(w) + 1 = k+1$, then ${\Cal U}^{w^{\prime}}$
is a twisted polynomial algebra of type ${\Cal U}^{w}_{\sigma,
D}[E_{\beta_{k+1}}]$, where the formulas for $\sigma$ and $D$ are
implicitly given in the formulas (2.2.1).
\endproclaim

\demo{Proof} (a) Using the fact that one can pass from one reduced
expression of $w$ to another by braid relations one reduces to the
case of rank $2$ where one repeats the analysis made by Lusztig
([L]).  (b) is clear by Lemma 2.2. \ \ \ $\square$

The elements $K_{\alpha}$ clearly normalize the algebras ${\Cal
U}^{w}$ and when we add them to these algebras we are performing an
iterated construction of Laurent twisted polynomials.  The resulting
algebras will be called ${\Cal B}^{w}$.

Since the algebras ${\Cal U}^{w}$ and ${\Cal B}^{w}$ are iterated
twisted polynomial rings with relations of the type 1.6(a) we can
consider the associated quasipolynomial algebras, and we
will denote them by $\overline{\Cal U}^{w}$ and $\overline{\Cal
B}^{w}$.  Notice that the latter algebras depend on the reduced
expression chosen for $w$.  Of course the defining relations for
these algebras are obtained from (2.2.1) by replacing the right-hand
side by zero.  We could of course also perform the same construction
with the negative roots but this is not strictly necessary since we
can simply apply the anti-automorphism $\kappa$ to define the
analogous negative objects.

\enddemo\subheading{\S 3. Degrees of algebras ${\Cal
    U}^{w}_{\varepsilon}$ and ${\Cal B}^{w}_{\varepsilon}$}

{\bf 3.1.} We specialize now the previous sections to the case $q =
\varepsilon$, a primitive $\ell$--th root of $1$.  Assuming that
$\ell^{\prime} > \underset{i}\to{\max}\ d_{i}$, we may consider the
specialized algebras:

$$
{\Cal U}_{\varepsilon} = {\Cal U}/(q-\varepsilon), \
{\Cal U}^{w}_{\varepsilon} = {\Cal U}^{w}/(q-\varepsilon), \
{\Cal B}^{w}_{\varepsilon} = {\Cal B}^{w}/(q-\varepsilon), \text{etc.}
$$
We have
obvious subalgebra inclusions ${\Cal U}^{w}_{\varepsilon} \subset {\Cal
B}^{w}_{\varepsilon} \subset {\Cal U}_{\varepsilon}$.

First, let us recall and give a simple proof of the following crucial
fact [DK1]:

\proclaim{Proposition} Elements $E^{\ell}_{\alpha}\ (\alpha \in R)$
and $K^{\ell}_{\beta}\ (\beta \in P)$ lie in the center
$Z_{\varepsilon}$ of ${\Cal U}_{\varepsilon}$ if $\ell^{\prime} >
\underset{i,j}\to{\max} |a_{ij}|$ (for any generalized Cartan matrix
$(a_{ij})$).
\endproclaim

\demo{Proof} The only non--trivial thing to check is that
$[E^{\ell}_{i},E_{j}] = 0$ for $i \neq j$.  From the ``Serre
relations'' it is immediate that $(ad_{\sigma_{-\alpha_{i}}}
E_{i})^{\ell^{\prime}} E_{j} = 0$.  Due to Corollary 1.1, this can be
rewritten as

$$E^{\ell^{\prime}}_{i}E_{j} =
\varepsilon^{-\ell^{\prime}(\alpha_{i}|\alpha_{j})}
E_{j}E^{\ell^{\prime}}_{i},$$
proving the claim.\ \ \ \ \ \ $\square$
\enddemo

As has been already remarked, the algebras ${\Cal
U}^{w}_{\varepsilon}$ and ${\Cal B}^{w}_{\varepsilon}$ are iterated
twisted polynomial algebras with relations of the type 1.6(a).
Proposition 3.1 shows that they satisfy conditions 1.6(b) and (c).
Hence Theorem 1.6 implies

\proclaim{Corollary} Algebras ${\Cal U}^{w}_{\varepsilon}$ and
$\overline{{\Cal U}}^{w}_{\varepsilon}$ (resp. ${\Cal
B}^{w}_{\varepsilon}$ and $\overline{{\Cal B}}^{w}_{\varepsilon}$)
have the same degree.\ \   \ \ \ $\square$
\endproclaim

{\bf 3.2.} We proceed to calculate the degrees of algebras
$\overline{\Cal U}^{w}_{\varepsilon}$ and $\overline{\Cal
B}^{w}_{\varepsilon}$.  Recall that these algebras are, up to
inverting some variables, quasipolynomial algebras whose generators
satisfy relations of type $x_{i}x_{j} = b_{ij}x_{j}x_{i},\ i,j =
1,\ldots ,s$, where the elements $b_{ij}$ have the special form
$b_{ij} = \varepsilon^{c_{ij}}$, the $c_{ij}$ being entries of a
skew--symmetric integral $s \times s$ matrix $H$.  As we have shown
in [DKP2, Proposition 2.2], considering $H$ as the matrix of a linear
map ${\Bbb Z}^{s} @>>> ({\Bbb Z}/(\ell ))^{s}$, the degree of the
corresponding twisted polynomial algebra is $\sqrt{h}$, where $h$ is
the number of elements of the image of this map.

Fix $w \in W$ and its reduced expression $w = s_{i_{1}}\ldots
s_{i_{k}}$.  We shall denote the matrix $H$ for the algebras
$\overline{\Cal U}^{w}_{\varepsilon}$ and $\overline{\Cal
B}^{w}_{\varepsilon}$ by $A$ and $S$ respectively.  First we describe
explicitly these matrices.

Let $d = 2$ unless $(a_{ij})$ is of type $G_{2}$ in which case $d =
6$, and let ${\Bbb Z}^{\prime} = {\Bbb Z}[d^{-1}]$.  Consider the
roots $\beta_{1},\ldots ,\beta_{k}$ as in Section 2.2, and consider
the free ${\Bbb Z}^{\prime}$--module $V$ with basis $u_{1},\ldots ,u_{k}$.
Define on $V$ a skew--symmetric
bilinear form by

$$\langle u_{i}|u_{j}\rangle = (\beta_{i}|\beta_{j})\ \text{if}\ i <
j.$$
Then $A$ is the matrix of this bilinear form in the basis $\{
u_{i}\}$.  Identifying $V$ with its dual $V^{*}$ using the given
basis, we may think of $A$ as a linear operator from $V$ to itself.

Furthermore,

$$S = \pmatrix A & -^{t}C \\ C & 0 \endpmatrix,$$ where $C$ is
the $n \times k$ matrix $((\omega_{i}|\beta_{j}))_{1 \leq i \leq
  n,1\leq j \leq k}.$ We may think of the matrix $C$ as a linear
map from the module $V$ with the basis $u_{1},\ldots ,u_{k}$ to
the module $Q^{\vee} \otimes_{\Bbb Z} {\Bbb Z}^{\prime}$ with the
basis $\alpha^{\vee}_{1},\ldots ,\alpha^{\vee}_{n}$.  Then we
have:

$$C(u_{i}) = \beta_{i},\ i = 1,\ldots ,k. \tag{3.2.1}$$

To study the matrices $A$ and $S$ we need the following

\proclaim{Lemma} Given $\omega = \sum^{n}_{i=1} \delta_{i} \omega_{i}$ with
$\delta_{i} = 0$ or $1$, set

$$I_{\omega} = \{ t \in \{ 1,\ldots ,k\}|s_{i_{t}}(\omega) \neq
\omega \}.$$
Then $\omega-w(\omega)=\sum_{t\in I_{\omega}}\beta_t$.
\endproclaim

\demo\nofrills{Proof}\ is by induction on the length of $w$.  Write $w =
w^{\prime}s_{i_{k}}$.  If $k \notin I_{\omega}$ then $w(\omega ) =
w^{\prime}(\omega)$ and we are done by induction.  Otherwise
$w(\omega ) = w^{\prime}(\omega - \alpha_{i_{k}}) = w^{\prime}(\omega
) - \beta_{k}$ and again we are done by induction.\  \ \ $\square$
\enddemo

Note that $\{ 1,2,\ldots ,k\} = \coprod^{n}_{i=1}
I_{\omega_{i}}.$

{\bf 3.3.}  Consider the operators: $M = (A\  -^{t}\!\!C)$ and $N = (C\ 0)$
so that $S = M \oplus N$.

\proclaim{Lemma}
(a) The operator $M$ is surjective.

(b) The vectors $v_{\omega} := (\sum_{t \in I_{w}} u_{t}) - \omega -
w(\omega )$, as $\omega$ runs through the fundamental weights, form a
basis of the kernel of $M$.

(c) $N(v_{\omega}) = \omega - w(\omega ) = \sum_{t \in I_{\omega}}
\beta_{t}.$
\endproclaim

\demo{Proof} (a) We have by a straightforward computation:

$$
S(u_{i} + \beta_{i}) =
A(u_{i})+B(u_{i})-^tB(\beta_i)=
\sum_{j<i}(\beta_j|\beta_i)u_j-\sum_{j>i}(\beta_j|\beta_i)u_j
+\beta_{i}-\sum_j(\beta_i|\beta_j)u_j
=$$
$$
-(\beta_{i}|\beta_{i})u_{i} - 2 \sum_{j >
i} (\beta_{i}|\beta_{j})u_{j} +\beta_{i},\; \text{and}\;
M(u_{i}+\beta_{i}) = -(\beta_{i}|\beta_{i})u_{i} - 2 \sum_{j>i}
(\beta_{i}|\beta_{j})u_{j}.
$$
Since $(\beta_{i}|\beta_{i})$ is invertible in ${\Bbb Z}^{\prime}$
the claim follows.

(b) Since the $n$ vectors $v_{\omega}$ are part of a basis and, by (a),
the kernel of $M$ is a direct summand of rank $n$, it is enough to show
that these vectors lie in the kernel.  To check that
$M(v_{\omega_{i}}) = 0$ is equivalent to seeing that $v_{\omega_{i}}$
lies in the kernel of the corresponding skew--symmetric form, i.e.
$\langle u_{j}|v_{\omega_{i}}\rangle = 0$ for all $j = 1,\ldots ,k$.
Recall that:
$$
\langle u_{i}|u_{j}\rangle = (\beta_{i}|\beta_{j})\;
\text{if}\; i < j, \langle
u_{j}|\beta\rangle = -(\beta_{j}|\beta),\ \beta\in P.
$$
Using Lemma 3.2, $v_{\omega} = \sum_{t \in I_{\omega}} (u_{t}-\beta_t) -
2w(\omega )$
and we have
$$
\langle u_{j}|v_{\omega_{i}}\rangle =\sum_{t\in I_{\omega_i}}
\langle u_{j}|u_{t}-\beta_t \rangle
-2\langle u_{j}|
w(\omega )\rangle
= 2
\sum_{t>j} (\beta_{j}|\beta_{t}) + 2(\beta_{j}|w(\omega_{i})) + a_{j},
\tag{3.3.1}
$$
where $a_{j} = 0$ if $j \notin I_{\omega_{i}}$ and $a_{j} =
(\beta_{j}|\beta_{j})$ otherwise.

We proceed by induction on $k = l(w)$.  Let us write
$v_{\omega_{i}}(w)$ to stress the dependence on $w$.  For $k = 0$
there is nothing to prove.  Let $w = w^{\prime}s_{i_{k}}$ with
$l(w^{\prime}) = l(w) - 1$.  We distinguish two cases according to
whether $i = i_{k}$ or not.

Case 1)  $i \neq i_{k}$, i.e. $k \notin
I_{\omega_{i}}$.

We have $I_{\omega_{i}}(w)=I_{\omega_{i}}(w')$ and
$w(\omega_i)=w'(\omega_i)$ so
that $v_{\omega_{i}}(w)
= v_{\omega_{i}}(w^{\prime})$ hence the claim follows by induction if
$j < k$.  For $j = k$ we obtain from (3.3.1):

$$\langle u_{k}|v_{\omega_{i}}\rangle = 2(\beta_{k}|w(\omega_{i})) =
2(w^{\prime}(\alpha_{i_{k}})|w^{\prime}(\omega_{i})) =
2(\alpha_{i_{k}}|\omega_{i}) = 0.$$

Case 2) $i_{k} = i$ so that $w = w^{\prime}s_{i}$ and
$w(\omega_i)=w's_i(\omega_i)=w'(\omega_i)-w'(\alpha_i)=
w'(\omega_i)-\beta_k$.

Then $v_{\omega_{i}}(w) = v_{\omega_{i}}(w^{\prime}) + u_{k} +
\beta_{k}$.
 For $j < k$ by induction we get:

$$\langle u_{j}|v_{\omega_{i}}\rangle = \langle u_{j}|u_{k}\rangle +
\langle u_{j}|\beta_{k}\rangle = (\beta_{j}|\beta_{k}) -
(\beta_{j}|\beta_{k}) = 0.$$
Finally if $j = k$ we have:

$$2(\beta_{k}|w(\omega_{i})) + (\beta_{k}|\beta_{k}) =
2(w^{\prime}\alpha_{i}|w^{\prime}(\omega_{i} - \alpha_{i}))
+(\alpha_{i}|\alpha_{i}) = 2(\alpha_{i}|\omega_{i}) -
(\alpha_{i}|\alpha_{i}) = 0.$$

(c) Using (3.1.2 ), we have: $N(v_{\omega}) = \sum_{t \in
I_{\omega}} \beta_{t}=\omega-w(\omega)$,  from Lemma
3.2.\ \ \  \ $\square$
\enddemo

\vskip10pt

{\bf 3.4.} In order to compute the kernel of $S$ we need to compute
the kernel of $N$ on the submodule spanned by the vectors
$v_{\omega_{i}}$.  Let us identify this module with the weight
lattice $P$ by identifying $v_{\omega_{i}}$ with $\omega_{i}$.  By
Lemma 3.3(c), we see that $N$ is identified with the map $1-w:\ P
@>>> Q$.  At this point we need the following fact:

\proclaim{Lemma} Let $\theta = \sum^{n}_{i=1} a_{i}\alpha_{i}$ be the
highest root of the root system $R$.  Let ${\Bbb Z}^{\prime\prime} =
{\Bbb Z}^{\prime\prime}[a^{-1}_{1},\ldots ,a^{-1}_{n}]$, and let
$M^{\prime} = M \otimes_{\Bbb Z} {\Bbb Z}^{\prime},\ M^{\prime\prime}
= M \otimes_{\Bbb Z} {\Bbb Z}^{\prime\prime}$ for $M = P$ or $Q$.
Then for any $w \in W$, the ${\Bbb Z}^{\prime\prime}$--submodule
$(1-w)P^{\prime\prime}$ of $Q^{\prime\prime}$ is a direct summand.
\endproclaim

\demo{Proof} Recall that one can represent $w$ in the form $w =
s_{\gamma_{1}} \ldots s_{\gamma_{m}}$ where $\gamma_{1},\ldots
,\gamma_{m}$ is a linearly independent set of roots (see e.g. [C]).
Since in the decomposition $\gamma^{\vee} = \sum_{i} r_{i}
\alpha^{\vee}_{i}$ one of the $r_{i}$ is $1$ or $2$, it follows that
$(1-s_{\gamma})P^{\prime} = {\Bbb Z}^{\prime}\gamma$.  Since $1-w =
(1-s_{\gamma_{1}} \ldots s_{\gamma_{m-1}})s_{\gamma_{m}} +
(1-s_{\gamma_{m}})$, we deduce by induction that

$$(1-w)P^{\prime} = \sum^{m}_{i=1} {\Bbb Z}^{\prime} \gamma_{m}
\tag{3.4.1}$$
Recall now that any sublattice of $Q$ spanned over ${\Bbb Z}$ by some
roots is a ${\Bbb Z}$--span of a set of roots obtained from $\Pi$ by
iterating the following procedure: \ add a highest root to the set of
simple roots, then remove several other roots from this set.  The
index of the lattice $M$ thus obtained in $M \otimes_{\Bbb Z} {\Bbb
Q} \cap Q$ is equal to the product of coefficients of removed roots
in the added highest root.  Hence it follows from (3.4.1) that

$$((1-w)P^{\prime\prime}) \otimes_{\Bbb Z} {\Bbb Q}\cap
Q^{\prime\prime} = (1-w)P^{\prime\prime},$$
proving the claim.\ \ \ \ $\square$
\enddemo

We call $\ell > 1$ a {\it good} integer if it is relatively prime
to $d$ and to all the $a_{i}$.

\proclaim{Theorem} If $\ell$ is a good integer, then

$$\deg {\Cal B}^{w}_{\varepsilon} = \deg \overline{\Cal
B}^{w}_{\varepsilon} = \ell^{\frac{1}{2}(\ell (w)+\text{rank}\
(1-w))}.$$
\endproclaim

\demo{Proof} From the above discussion we see that $\deg \overline{\Cal
B}^{w}_{\varepsilon} = \ell^{s}$, where $s = (\ell
(w)+n)-(n-\text{rank}(1-w))$, which together with Corollary 3.1 proves the
claim.\ \ \ $\square$
\enddemo

{\bf 3.5.} We pass now to ${\Cal U}^{w}_{\varepsilon}$.  For this we
need to compute the image of the matrix $A$.  Computing first its
kernel, we have that $Ker A$ is identified with the set of linear
combinations $\sum_{i} c_{i}v_{\omega_{i}}$ for which $\sum_{i}
c_{i}(\omega_{i} + w(\omega_{i})) = 0$ i.e. $\sum_{i} c_{i}
\omega_{i} \in \ker (1+w)$.  This requires a case by case analysis.
A simple case is when $w_{0} = -1$, so that $1 + w = w_{0}(-1 +
w_{0}w)$ and one reduces to the previous case.  Thus we get

\proclaim{Proposition} If $w_{0} = -1$ (i.e. for types different from
$A_{n},\ D_{2n+1}$ and $E_{6}$) and if $\ell$ is a good integer, we
have:

$$\deg {\Cal U}^{w}_{\varepsilon} = \deg \overline{\Cal
U}^{w}_{\varepsilon} = \ell^{\frac{1}{2}(\ell (w) +\text{rank}(1+w)-n)}.$$
\endproclaim

Let us note the special case $w = w_{0}$.  Remark that defining
$^{t}\omega := -w_{0}(\omega)$ we have an involution $\omega @>>> \
^{t}\omega$ on the set of fundamental weights.  Let us denote by $s$
the number of orbits of this involution.

\proclaim{Theorem} If $\varepsilon$ is a primitive $\ell$--th root of $1$,
where $\ell$ is an integer greater than $1$ and relatively prime to $d$, then
the algebras ${\Cal U}_{\varepsilon}^{w_{0}}$ and ${\Cal
B}_{\varepsilon}^{w_{0}}$
have degrees $\ell^{\frac{N-s}{2}}$ and $\ell^{\frac{N+s}{2}}$ respectively.
\endproclaim

\demo{Proof} In this case $l(w_{0}) = N$ and the maps $\omega @>>> \omega +
w_{0}(\omega )$ and $\omega @>>> \omega - w_{0}(\omega )$ are $\omega @>>>
\omega - \ ^{t}\omega$ and $\omega @>>> \omega + \ ^{t}\omega$ and so their
ranks are clearly $n-s$ and $s$ respectively. $\square$
\enddemo

\subheading{\S 4. Poisson structure}

{\bf 4.1.} Before we revert to the discussion of our algebras we want to make
a general remark.  Assume that we have a manifold $M$ and a vector
bundle $V$ of algebras with 1 (i.e., $1$ and the multiplication map are
smooth sections).  We identify the functions on $M$ with the sections of $V$
which are multiples of $1$.  Let $D$ be a derivation of $V$, i.e., a derivation
of the algebra of sections which maps the algebra of functions on $M$ into
itself and let $X$ be the corresponding vector field on $M$.

\proclaim{Proposition}  For each point $p \in M$ there exists a neighborhood
$U_{p}$ and a map $\varphi_{t}$ defined for $|t|$ sufficiently small on
$V|U_{p}$ which is a morphism of vector bundles covering the germ of the
$1$-parameter group generated by $X$ and is also an isomorphism of algebras.
\endproclaim

\demo{Proof} The hypotheses on $D$ imply that it is a vector field on $V$
linear on the fibers, hence we have the existence of a local lift of the
$1$-parameter group as a morphism of vector bundles.  The condition of being
a derivation implies that the lift preserves the multiplication section i.e.
it is a morphism of algebras.\ \ \ $\square$
\enddemo

We will have to consider a variation of this: suppose $M$ is a Poisson
manifold and assume furthermore that the Poisson structure lifts to $V$ i.e.
for each (local) function $f$ and section $s$ we have a Poisson bracket which
is a derivation.  This means that we have a lift of the Hamiltonian vector
fields as in the previous proposition.  We deduce:

\proclaim{Corollary} Under the previous hypotheses, the fibers of $V$ over
points of a given symplectic leaf of $M$ are all isomorphic as algebras.
\endproclaim

\demo{Proof} The proposition implies that in a neighborhood of a point in a
leaf the algebras are isomorphic but since the notion of isomorphism is
transitive this implies the claim.\ \ \ $\square$
\enddemo

{\bf 4.2.}  Let us recall some basic facts on Poisson groups (we refer to [D],
[STS], [LW]
for basic definitions and properties).  Since a Poisson structure on a
manifold $M$ is a special type of a section of $\Lambda^{2}T(M)$ it can be
viewed as a linear map from $T^{*}(M)$ to $T(M)$.  The image of this map is
thus a distribution on the manifold $M$. It can be integrated so that we have
a decomposition of $M$ into symplectic leaves which are connected locally
closed
submanifolds whose tangent spaces are the spaces of the distribution.  In fact
in our case the leaves will turn out to be Zariski open sets of algebraic
subvarieties.

For a group $H$ the tangent space at each point can be identified to the Lie
algebra ${\frak h}$ by left translation and thus a Poisson structure on $H$
can be given as a family of maps $\gamma_{h}:\ {\frak h}^{*} @>>> {\frak h}$
as $h \in H$.  Let $G$ be an algebraic group and $H,K \subset G$ algebraic
subgroups.  We shall say that $(G,H,K)$ form a Manin triple of algebraic
groups if their corresponding Lie algebras $({\frak g},{\frak h},{\frak k})$
form a Manin triple, i.e., (cf. [D], [LW]) if ${\frak g}$ has a
non-degenerate,
symmetric invariant bilinear form with respect to which the Lie subalgebras
${\frak h}$
and ${\frak k}$ are isotropic and ${\frak g} = {\frak h} \oplus {\frak k}$ (as
vector spaces).  Then it follows that we have a canonical isomorphism ${\frak
h}^{*} = {\frak k}$.  Having identified ${\frak h}^{*}$ with ${\frak k}$, the
Poisson structure on $H$ is thus described by giving for every $h \in H$ a
linear map $\gamma_{h}:\ {\frak k} @>>> {\frak h}$.

Let $x \in {\frak k}$, consider $x$ as an element of ${\frak g}$, set $\pi :
{\frak g} @>>> {\frak h}$ to be the projection with kernel ${\frak k}$.
Set:

$$\gamma_{h}(x) = (Ad h)\pi(Ad h)^{-1}(x).$$
Then one can verify (as in [LW]) that the corresponding tensor satisfies the
required properties of a Poisson structure.  (In fact any Poisson structure on
$H$ can be obtained in this way.)

Notice now that the (restriction of the) canonical map:

$$\delta :\ H @>>> G/K$$
is an \'{e}tale covering of some open set in $G/K$.  Thus for every point $h
\in H$ we can identify the tangent space to $H$ in $h$ with the tangent space
to $G/K$ at $\delta (h)$.  By using right translation we can then
identify the tangent space to $G/K$ at $\delta (h)$ with ${\frak g}
/(Ad h){\frak k}$, the tangent space at $h \in H$ with ${\frak h}$ by right
translation and the isomorphism between them with the projection ${\frak h}
@>>> {\frak g}/(Ad h){\frak k}$.

Using all these identifications one verifies that the map $\gamma_{h}$
previously considered is the map induced by differentiating the left
$K$-action on $G/K$.  From this it follows:

\proclaim{Proposition} The symplectic leaves for the symplectic structure on
$H$ coincide with the connected components of the preimages under $\delta$ of
$K$-orbits under the left multiplications on $G/K$.
\endproclaim

Consider now a quotient Poisson group $S$ of $H$, that is $S$ is a quotient
group of $H$ and the ring ${\Bbb C}[S] \subset {\Bbb C}[H]$ is a Poisson
subalgebra.  Let $U$ be the kernel of the quotient homomorphism $\varphi :\ H
@>>> S$, let ${\frak s} = Lie\ S,\ {\frak u} = Lie \ U$ and $d\varphi :\
{\frak h} @>>> {\frak s}$ the Lie algebra quotient map.  Then ${\frak u}$ is
an ideal
in ${\frak h}$ and we identify ${\frak s}^{*}$ with a subspace of ${\frak
h}^{*} = {\frak k}$ by taking ${\frak u}^{\perp} \subset {\frak g}$ under the
invariant form and intersecting it with ${\frak k}$.  Then for $p \in S$ the
linear map: ${\overline \gamma}_{p}:\ {\frak s}^{*} @>>> {\frak s}$ giving
rise to the Poisson structure is given by:

$${\overline \gamma}_{p} = (d\varphi)\cdot (\gamma_{\tilde p}|_{{\frak
s}^{*}})$$
where ${\tilde p} \in H$ is any representative of $p$ $({\overline
\gamma}_{p}$ is
independent of the choice of ${\tilde p})$.

The construction of the Manin triple corresponding to the Poisson manifold $S$
is obtained from the following simple fact:

\proclaim{Lemma} Let $({\frak g},{\frak h},{\frak k})$ be a Manin triple of
Lie algebras, and let ${\frak u} \subset {\frak h}$ be an ideal such that
${\frak u}^{\perp}$ (in ${\frak g}$) intersected with ${\frak k}$ is a
subalgebra of the Lie algebra ${\frak k}$.  Then

(a) ${\frak u}^{\perp}$ is a subalgebra of ${\frak g}$ and ${\frak u}$ is an
ideal of ${\frak u}^{\perp}$.

(b) $({\frak u}^{\perp}/{\frak u},\ {\frak h}/{\frak u},\ {\frak k} \cap
{\frak u}^{\perp})$ is a Manin triple, where the bilinear form on ${\frak
u}^{\perp}$ is induced by that on ${\frak g}$.
\endproclaim

\demo\nofrills{Proof}\ is straightforward.\ \ \ \ $\square$
\enddemo

{\bf 4.3.} In the remaining sections we will apply the above remarks
to the Poisson groups associated to the Hopf
algebra ${\Cal U}_{\varepsilon}$ and its Hopf subalgebra ${\Cal
B}_{\varepsilon} := {\Cal B}^{w_{0}}_{\varepsilon}$, and will derive
some results on representations of the algebra ${\Cal
B}_{\varepsilon}$.  From now on $\varepsilon$ is a primitive
$\ell$-th root of $1$ where $\ell > 1$ is relatively prime to $d$.

Let $Z_{0}$ (resp. $Z^{+}_{0}$) be the subalgebra of ${\Cal
U}_{\varepsilon}$ (resp. ${\Cal B}_{\varepsilon}$) generated by the
elements $E^{\ell}_{\alpha}$ with $\alpha \in R$ (resp. $\alpha \in
R^{+}$) and $K_{\beta}$ with $\beta \in P$.  (We assume fixed $a$
reduced expression of $w_{0}$; $Z_{0}$ and $Z^{+}_{0}$ are
independent of this choice [DK1].)  Recall that they are central
subalgebras (Proposition 3.1).

It was shown in [DK1] that $Z_{0}$ and $Z^{+}_{0}$ are Hopf
subalgebras, hence $\text{Spec}\ Z_{0}$ and $\text{Spec}\ Z^{+}_{0}$
have a canonical structure of an affine algebraic group.
Furthermore, since ${\Cal U}_{\varepsilon}$ is a specialization of
the algebra ${\Cal U}$ at $q = \varepsilon$, the center
$Z_{\varepsilon}$ of ${\Cal U}_{\varepsilon}$ posesses a canonical
Poisson bracket given by the formula:

$$\{ a,b\} = \frac{[{\hat a},{\hat b}]}{2\ell^{2}(q-\varepsilon)} \mod
(q-\varepsilon ),\ \ a,b \in Z_{\varepsilon},$$
where ${\hat a}$ denotes the preimage of $a$ under the canonical
homomorphism ${\Cal U} @>>> {\Cal U}_{\varepsilon}$.  The algebras
$Z_{0}$ and $Z^{+}_{0}$ are Poisson subalgebras of $Z_{\varepsilon}$,
thus $\text{Spec}\ Z_{0}$ and $\text{Spec}\ Z^{+}_{0}$ have a
canonical structure of Poisson algebraic groups, $\text{Spec}\
Z^{+}_{0}$ being a quotient Poisson group of $\text{Spec}\ Z_{0}$.

In [DKP1] an explicit isomorphism was constructed between the Poisson
group $\text{Spec}\ Z_{0}$ and a Poisson group $H$ which is described
below.  We shall identify these Poisson groups.

Let $G$ be the connected simply connected algebraic group associated
to the Cartan matrix $(a_{ij})$ and let ${\frak g}$ be its (complex)
Lie algebra.  We fix the triangular decomposition ${\frak g} = {\frak
u}_{-} + {\frak t} + {\frak u}_{+}$, let ${\frak b}_{\pm} = {\frak t}
+ {\frak u}_{\pm}$, and denote by $(.|.)$ the invariant bilinear form
on ${\frak g}$ which on the set of roots $R \subset {\frak t}^{*}$
coincides with that defined in Section 2.1.  Let $U_{\pm},\
B_{\pm}$ and $T$ be the algebraic subgroups of $G$ corresponding to
Lie algebras ${\frak u}_{\pm},\ {\frak b}_{\pm}$ and ${\frak t}$.
Then as an algebraic group, $H$ is the following subgroup of $G
\times G$:

$$H = \{ (tu_{+},t^{-1}u_{-})|t \in T,\ u_{\pm} \in {U}_{\pm}\}.$$
The Poisson structure on $H$ is given by the Manin triple $({\frak g}
\oplus {\frak g},\ {\frak h},\ {\frak k})$, where

$$\align
&{\frak h} = \{ (t + u_{+},\ -t + u_{-})|t \in {\frak t},u_{\pm} \in
{\frak u}_{\pm}\} ,\\
&{\frak k} = \{ (g,g)|g \in {\frak g}\},
\endalign$$ and the invariant bilinear form on ${\frak g} \oplus
{\frak g}$ is

$$((x_{1},x_{2})|(y_{1},y_{2})) = -(x_{1}|y_{1}) + (x_{2}|y_{2}).$$

We identify the group $B_{+} = H/\{ (1,u_{-})|u_{-} \in U_{-}\}.$
The Manin triple generating Poisson structure on $B_{+}$ is obtained
from $({\frak g} \oplus {\frak g},{\frak h},{\frak k})$ by taking the
ideal ${\frak u} = \{ (0,u_{-}),u_{-} \in {\frak u}_{-}\}$ and applying the
construction given by Lemma 4.2.  We clearly obtain the triple
$({\frak g} \oplus {\frak t}, {\frak b}_{+},{\frak b}_{-})$, where we used
identifications

$${\frak b}_{\pm} = \{ (u_{\pm}-t,\pm t)|u_{\pm} \in {\frak u}_{\pm},t
\in {\frak t}\}.$$

According to the general recipe of Proposition 4.2, the symplectic
leaves of the Poisson group $B_{+}$ are obtained as follows.  We
identify the groups $B_{\pm}$ with the following subgroups of $G
\times T$:

$$B_{\pm} = \{ (t^{-1}u_{\pm},t^{\pm 1})|t \in T, u_{\pm} \in U_{\pm}\}.$$
The inclusion $B_{+} \subset G \times T$ induces an etale morphism

$$\delta :\ B_{+} @>>> (G \times T)/B_{-}.$$
Then the symplectic leaves of $B_{+}$ are the connected components of
the preimages under the map $\delta$ of $B_{-}$-orbits on $G \times
T/B_{-}$ under the left multiplication.

In order to analize the $B_{-}$-orbits on $G \times T/B_{-}$, let
$\mu_{\pm}:\ B_{\pm} @>>> T$ denote the canonical homomorphisms with
kernels $U_{\pm}$ and consider the equivariant isomorphism of
$B_{-}$-varieties $\gamma :\ G/U_{-} @>>> (G \times T)/B_{-}$ given
by $\gamma (g{\Cal U}_{-}) = (g,1)B_{-}$, where $B_{-}$ acts on
$G/U_{-}$ by

$$b(gU_{-}) = bg\mu_{-}(b)U_{-}. \tag{4.3.1}$$
Then the map $\delta$ gets identified with the map  $\delta :\ B_{+}
@>>> G/U_{-}$ given by

$$\delta (b) = b\mu_{+}(b) U_{-}.$$

We want to study the orbits of the action (4.3.1) of $B_{-}$ on
$G/U_{-}$.  Consider the action of $B_{-}$ on $G/B_{-}$ by
left multiplication.  Then the canonical map  $\pi : G/U_{-}
@>>> G/B_{-}$ is $B_{-}$-equivariant, hence $\pi$ maps every
$B_{-}$-orbit ${\Cal O}$ in $G/U_{-}$ to a $B_{-}$-orbit in
$G/B_{-}$, i.e., a Schubert cell $C_{w} = B_{-}wB_{-}/B_{-}$ for some
$w \in W$.  We shall say that the orbit ${\Cal O}$ is associated to
$w$.

\vskip 5pt
\noindent REMARK.  We have a sequence of maps:

$$B_{+} @>\delta >> (G \times T)/B_{-}
@>\underset{\sim}\to{\gamma^{-1}} >> G/U_{-}
@>\pi >>  G/B_{+}.$$
Let $\psi = \pi \circ \gamma^{-1} \circ \delta$ and $X_{w} = B_{+}
\cap B_{-}wB_{-}$.  Then:

$$\pi^{-1}(C_{w}) = B_{-}wB_{-}/U_{-}\ \text{and}\ \psi^{-1}(C_{w})
= X_{w}.$$

We can prove now the following

\proclaim{Proposition} Let ${\Cal O}$ be a $B_{-}$-orbit in $G/U_{-}$ under
the action (4.3.1) associated to $w \in W$.  Then the
morphism:

$$\pi |_{\Cal O}:\ {\Cal O} @>>> C_{w}$$
is a principal torus bundle with structure group:

$$T^{w} := \{ w^{-1}(t)t^{-1},\ \text{where}\ t \in T\}.$$
In particular:

$$\dim {\Cal O} = \dim C_{w} + \dim T^{w} = l(w) +\ \text{rank}(I-w).$$
\endproclaim

\demo{Proof} For $g \in G$ we shall write $[g]$ for the coset
$gU_{-}$.  The morphism $\pi$ is clearly a principal $T$-bundle
with $T$ acting on the right by $[g]t := [gt]$.  The action
(4.3.1) of $B_{-}$ commutes with the right $T$-action so that $T$
permutes the $B_{-}$-orbits.  Each $B_{-}$-orbit is a principal
bundle whose structure group is the subtorus of $T$ which
stabilizes the orbit.  This subtorus is independent of the orbit
since $T$ is commutative.  In order to compute it we procede as
follows.  Let $[g_{1}],[g_{2}]$ be two elements in ${\Cal O}$
mapping to $w \in C_{w}$.  We may assume that $g_{1} = nh,\ g_{2}
= nk$ with $h,k \in T$ uniquely determined, where $n \in
N_{G}(T)$ is representative of $w$.  Suppose that $b[nh] =
b[nk]$, $b \in B_{-}$.  We can first reduce to the case $b = t
\in T$; indeed, writing $b = ut$ we see that $u$ must fix $w \in
C_{w}$ hence $un = nu^{\prime}$ with $u^{\prime} \in U_{-}$ and
hence $u$ acts trivially on $t[nh]$.  Next we have that, by
definition of the $T$-action (4.3.1),

$$[nk] = [tnht^{-1}] = [n(n^{-1}tnht^{-1})]$$
hence $k = n^{-1}tnht^{-1}$ or $k = h(h^{-1}n^{-1}tnht^{-1}) =
h(n^{-1}tnt^{-1})$ as required.\ \ \ $\square$
\enddemo

\proclaim{Lemma}  Let ${\Cal O} \subset B_{+}$ be a symplectic leaf
associated to $w$.  Then ${\Cal O}T = X_{w}$.
\endproclaim

\demo{Proof}  From our proof we know that the map $\delta$ is a
principal $T$-bundle and $T$ permutes transitively the leaves lying
over $C_{w}$.\ \ \ $\square$
\enddemo

We thus have a canonical stratification of $B_{+}$, indexed by the
Weyl group, by the subsets $X_{w}$.  Each such subset is a union of
leaves permuted transitively by the right multiplications of the
group $T$.

We say that a point $a \in \text{Spec}Z^{+}_{0} =
B_{+}$ lies over $w$ if $\psi (a) \in C_{w}$.

{\bf 4.4.}  Recall that $T = {\Bbb C}^{\times} \otimes_{\Bbb Z}
Q^{\vee}$ and therefore any $\lambda \in P = \text{Hom}_{\Bbb
Z}(Q^{\vee},{\Bbb C}^{\times})$ defines a homomorphism (again denoted
by) $\lambda :\ T @>>> {\Bbb C}^{\times}$.  For each $t \in T$ we
define an automorphism $\beta_{t}$ of the algebra ${\Cal
B}_{\varepsilon}$ by:

$$\beta_{t}(K_{\alpha}) = \alpha (t)K_{\alpha},\
\beta_{t}(E_{\alpha}) = \alpha (t)E_{\alpha}.$$
Note that the automorphisms $\beta_{t}$ leave $Z^{+}_{0}$ invariant
and permute transitively the leaves of each set $\psi^{-1}(C_{w})
\subset B_{+}$.

Given $a \in B_{+} = \text{Spec}\ Z^{+}_{0}$, denote by $m_{a}$ the
corresponding maximal ideal of $Z^{+}_{0}$ and let
$$A_{a} = \Cal B_{\varepsilon}/m_{a}{\Cal B}_{\epsilon}.$$
These are finite-dimensional algebras and we may also consider these
algebras as algebras with trace in order to use the techniques of
[DKP2].

\proclaim{Theorem} If $a,b \in \text{Spec}\ Z^{+}_{0}$ lie over the
same element $w \in W$, then the algebras $A_{a}$ and $A_{b}$ are
isomorphic (as algebras with trace).
\endproclaim

\demo{Proof}  We just apply Proposition 4.1 to the vector bundle of
algebras $A_{a}$ over a symplectic leaf and the group $T$ of algebra
automorphisms which permutes the leaves in $\psi^{-1}(C_{w})$
transitively.\ \ \ $\square$
\enddemo

{\bf 4.5.} Let $B^{w} := B_{+} \cap wB_{-}w^{-1}$ and $U^{w}
:= U_{+} \cap wU_{-}w^{-1}$ so that $B^{w} = U^{w}T.$  Set also $U_{w} :=
U_{+} \cap wU_{+}w^{-1}$.  One knows that $\dim B^{w} = n + l(w)$ and
that the multiplication map:

$$\sigma :\ U_{w} \times B^{w} @>>> B_{+}$$
is an isomorphism of algebraic varieties.  We define the map

$$p_{w} :\ B_{+} @>>> B^{w}$$
to be the inverse of $\sigma$ followed by the projection on the
second factor.

\proclaim{Proposition} The map

$$p_{w}|_{X_{w}} :\ X_{w} @>>> B^{w}$$
is birational.
\endproclaim

\demo{Proof} We need to exhibit a Zariski open set $\Omega \subset
B^{w}$ such that for any $b \in \Omega$ there is a unique $u \in
U_{w}$ with $ub \in X_{w}$.

Let $n \in N_{G}(T)$ be as above a representative for $w$ so that:

$$X_{w} = \{ b \in B_{+}|b = b_{1}nb_{2},\ \text{where}\ b_{1},b_{2}
\in B_{-}\}.$$
Consider the Bruhat cell $B_{-}n^{-1}B_{-} \subset G$.  Every element
in $B_{-}n^{-1}B_{-}$ can be written uniquely in the form:

$$bn^{-1}u,\ \text{where}\ b \in n^{-1}B^{w}n,\ u \in U_{-}.$$
The set $B_{+}U_{-} = B_{+}B_{-}$ is open dense and so it intersects
$B_{-}n^{-1}B_{-} = n^{-1}B^{w}U_{-}$ in a non-empty open set which
is clearly $B_{-}$-stable for the right multiplication, hence\newline
$B_{+}B_{-} \cap B_{-}n^{-1}B_{-} = n^{-1}\Omega U_{-}$ for some non
empty open set $\Omega \subset B^{w}$.  In particular $\Omega \subset
nB_{+}B_{-} = nU_{w}B^{w}U_{-}$.  Take $b \in \Omega$ and write it as
$b = nxcv$ with $x \in U_{w},\ c \in B^{w},\ v \in U_{-}$.  By the
remarks made above this decomposition is unique; furthermore,
$nxn^{-1} \in U_{w},\ ncn^{-1} \in B_{-}$.  For the element $n^{-1}b
= xcv$ we have by construction that $xcv \in B_{-}n^{-1}B_{-}$ and
$nx^{-1}n^{-1}b = (ncn^{-1})nv \in B_{-}nB_{-}$ and $nx^{-1}n^{-1}
\in U_{w}$.  Thus setting $u := nx^{-1}n^{-1}$ we have found $u \in
U_{w}$ such that $ub \in X_{w}$.  This $u$ is unique since the
element $x$ is unique.\ \ \ $\square$
\enddemo

We are ready now for the concluding theorem which is in the spirit of
the conjecture formulated in [DKP1].

\proclaim{Theorem}  Let $p \in X_{w}$ be a point over $w \in W$ and
let $A_{p}$ be the corresponding algebra.  Assume that $l$ is a good
integer.  Then the dimension of each irreducible representation of
$A_{p}$ is divisible by $l^{\frac{1}{2}(l(w)+\text{rank}(1-w))}$.
\endproclaim

\demo{Proof}  Consider the algebra $\Cal B^{w}_{\varepsilon}$ for which we
know by Theorem 3.5 that
$$\deg \Cal B^{w}_\varepsilon = l^{\frac{1}{2}(l(w) +
\text{rank}(1-w))}.$$
The subalgebra $Z_{0,w}$ of $Z_{0}$ generated by the elements
$K^{l}_{\lambda}$ and $E^{l}_{\alpha}$, where $\lambda \in P$ and
$\alpha \in R^{+}$ is such that $-w^{-1}\alpha \in R^{+}$, is isomorphic to
the coordinate ring of $B^{w}$, and ${\Cal B}^{w}_{\varepsilon}$ is a
finite free module over $Z_{0,w}$.  Thus by [DKP2] there is a non
empty open set ${\Cal A}$ of $B^{w}$ such that for $p \in {\Cal A}$
any irreducible representation of ${\Cal B}^{w}$ lying over $p$ is of
maximal dimension, equal to the degree of ${\Cal
B}^{w}_{\varepsilon}$.  Now the ideal $I$ defining $X_{w}$ has
intersection $0$ with $Z_{0,w}$ and so when we restrict a generic
representation of ${\Cal B}_{\varepsilon}$ lying over points of
$X_{w}$ to the algebra ${\Cal B}^{w}_{\varepsilon}$ we have, as a
central character of $Z_{0,w}$, a point in ${\Cal A}$. Thus the
irreducible representation restricted to $\Cal B^w_\varepsilon$ has all its
composition factors irreducible of dimension equal to $\deg \Cal
B^{w}_\varepsilon$.  This proves the claim.\ \ \ $\square$
\enddemo

It is possible that the dimension of any irreducible representation
of $\Cal B_{\varepsilon}$ whose central character restricted to
$Z^{+}_{0}$ is a point of $X_{w}$ is exactly $\ell^{\frac{1}{2}(\ell (w) +
\text{rank}(1-w))}$.  This fact if true would require a more detailed
analysis in the spirit of Section 1.3.

We would like, in conclusion, to propose a more general conjecture,
similar to one of the results of [WK] on solvable Lie algebras of
characteristic $p$.

Let $A$ be an algebra over ${\Bbb C}[q,q^{-1}]$ on generators
$x_{1},\ldots ,x_{n}$ satisfying the following relations:

$$x_{i}x_{j} = q^{h_{ij}}x_{j}x_{i} + P_{ij} \ \text{if}\ i > j,$$
where $(h_{ij})$ is a skew-symmetric matrix over ${\Bbb Z}$ and
$P_{ij} \in {\Bbb C}[q,q^{-1}][x_{1},\ldots ,x_{i-1}]$.  Let $\ell >
1$ be an integer relatively prime to all elementary divisors of the
matrix $(h_{ij})$ and let $\varepsilon$ be a primitive $\ell$-th
root of $1$.  Let $A_{\varepsilon} = A/(q-\varepsilon )$ and assume
that all elements $x^{\ell}_{i}$ are central.  Let $Z_{0} = {\Bbb
C}[x^{\ell}_{1},\ldots ,x^{\ell}_{n}]$; this algebra has a canonical
Poisson structure.

\noindent {\smc Conjecture.} Let $\pi$ be an irreducible
representation of the algebra $A_{\varepsilon}$ and let ${\Cal
  O}_{\pi} \subset \text{Spec}\ Z_{0}$ be the symplectic leaf
containing the restriction of the central character of $\pi$ to
$Z_{0}$.  Then the dimension of this representation is equal to
$\ell^{\frac{1}{2}\dim {\Cal O}_{\pi}}$.

This conjecture of course holds if all $P_{ij}$ are $0$, and it is in
complete agreement with Theorems 1.6, 3.5 and 4.5.

\enddocument